\documentclass[prb,twocolumn,showpacs]{revtex4}
\usepackage{epsfig,color}

\begin{document}
\title{Electron transfer through a multiterminal quantum ring: \\ magnetic forces and elastic scattering
effects}
\author{B. Szafran} \affiliation{Faculty of Physics and Applied
Computer Science, AGH University of Science and Technology, al.
Mickiewicza 30, 30-059 Krak\'ow, Poland}
\author{M.R. Poniedzia{\l}ek } \affiliation{Faculty of  Physics and Applied
Computer Science, AGH University of Science and Technology, al.
Mickiewicza 30, 30-059 Krak\'ow, Poland}

\date{\today}

\begin{abstract}
We study electron transport through a semiconductor quantum ring with one input and two output terminals
for an elastic scatterer present within one of the arms of the ring. We demonstrate that the scatterer not
only introduces asymmetry in the transport probability to the two output leads but also reduces the visibility of the Aharonov-Bohm conductance oscillations.
This reduction occurs in spite of the phase coherence of the elastic scattering and is due to interruption of the electron circulation
around the ring by the potential defect.
The results are in a qualitative agreement with a recent experiment by Strambini et al. [Phys. Rev. B {\bf 79}, 195443 (2009)]. We also indicate that the magnetic symmetry of the sum of conductance of both the output leads as obtained in the experiment can be understood as
resulting from the invariance of backscattering to the input lead with respect to the magnetic field orientation.
\end{abstract}
 \maketitle

\section{Introduction}

Although studies of electron transport in semiconductor rings
have a long history,\cite{but,tip} the interest in this field is sustained by progress of experimental techniques.
In particular, transport through quantum rings containing a number of confined electrons was realized within the last decade\cite{firer,keyser}
and double concentric quantum rings were recently studied.\cite{muhle} Moreover, self-interference of electrons
injected individually into the quantum ring was observed with a time-resolved technique.\cite{gustav} The Fermi level wave functions were probed by conductance measurements for the ring potential landscape
perturbed by a tip of atomic force microscope.\cite{sgs} The effect of magnetic forces on quantum ring conductance was
studied experimentally in Ref. [\onlinecite{wlochy}].

In presence of the magnetic forces the electron wave function enters
both arms of the quantum ring with an unequal amplitude.\cite{time}
For two-terminal rings the preferential injection of the electron wave function
 into one of the arms of the ring leads to attenuation of the Aharonov-Bohm oscillation at high magnetic field.\cite{time}
It was demonstrated\cite{epl} that for rings with three terminals
at high field in addition to the vanishing oscillation amplitude
the magnetic forces produce  a distinct imbalance of the electron transport probabilities to the two output leads.
Both the high field reduction of the oscillation amplitude and the imbalance in the conductance of the two output leads
were indeed found in the recent experiment.\cite{wlochy}
However,
the experimental data \cite{wlochy} differ from the theoretical results \cite{epl} within the range of weak magnetic fields, namely: i) the measured conductance of one of the output leads significantly exceeds the other near $B=0$ (Fig. 1 of Ref. [\onlinecite{wlochy}]) and ii) already for low magnetic fields the experimental Aharonov-Bohm conductance oscillations have a low amplitude. The first feature suggests that the potential
landscape within the ring is asymmetric
and the second was attributed\cite{wlochy} to decoherence.
The estimated\cite{wlochy}  coherence length is 320 nm, which is surprisingly short -- an order of magnitude shorter than the estimate
for the two-dimensional electron gas\cite{dec} for the temperature of 350 mK applied in the experiment.\cite{wlochy}
In the present paper we indicate that the observed features of the conductance can also be explained for purely coherent transport as resulting from
the elastic scattering effects which do not randomize the phase but reduce the circulation of the electron around the arms of the ring.
We perform a systematic study of the electron transport in a three-terminal ring containing a potential defect.
We find that only a repulsive and not an attractive scattering center may explain the conductance
features as seen in the experiment.

The sum of conductance of both the output leads turns out \cite{wlochy} to be an even
function of the magnetic field, which is reminiscent of the Onsager symmetry for two-terminal devices.\cite{onsager}
The sum of the transfer probabilities to the left $T_l$ and right $T_r$ output leads $T=T_l+T_r$ as found in the
simulation of symmetric rings\cite{epl} is also an even function of $B$, but for evident kinetic reasons which no longer hold for asymmetric rings.
We demonstrate below that for the ring with a defect the kinetics of the electron transfer and the electron trajectory are very different
for opposite magnetic field orientations and that the observed\cite{wlochy} $T(B)=T(-B)$ symmetry is due
to the invariance of the backscattering with respect to the orientation of the $B$ vector.

Multiterminal rings constitute basic elements for construction of arrays, which are
used in detection of the Aharonov-Casher effect\cite{acshift} and are attractive for construction of programable quantum gates.\cite{peeters}
The three-terminal quantum rings were investigated in the context of the Kondo density of states.\cite{lt}
It was demonstrated that the spin-orbit coupling effects in three terminal rings can be used for construction of
electron spin beam splitters.\cite{bsp}

The current as carried
by a steady electron flow at the Fermi level can be determined from the Hamiltonian eigenequation.
However, for the purpose of the present study we choose to
employ a time dependent approach providing a clear picture of the
electron trajectory which in the present problem appears as quite complex. With the wave packet description of the electron motion one can approach the time independent
monoenergetic limit arbitrarily close.

\begin{figure}[ht!]
\epsfysize=150mm \epsfbox[100 280 300 834]{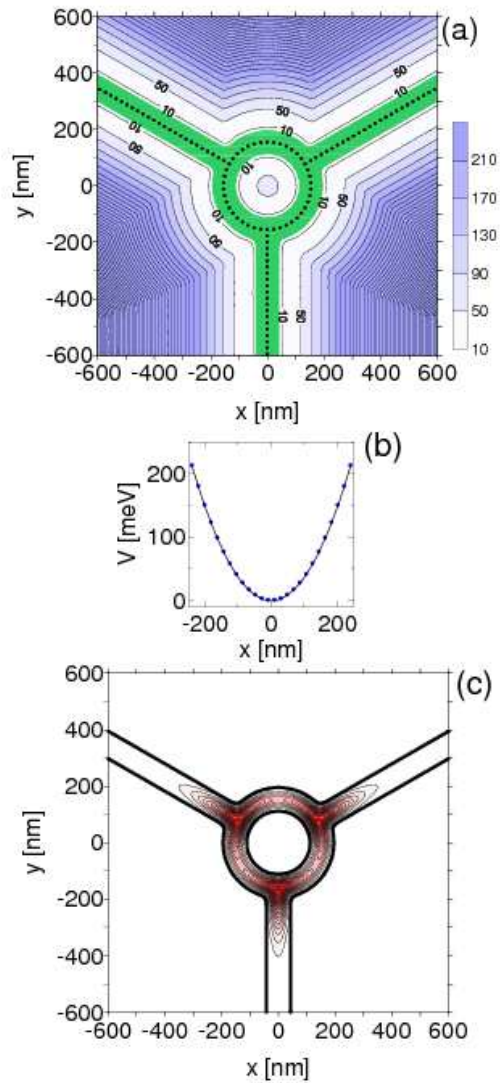} \\
\caption{(a) The model of the three-terminal quantum ring. The dots
indicate the positions of the centers of the Gaussian basis (\ref{gf}). The green area
shows the estimated confinement region accessible to the traveling electron (see text)
for the width of the channel equal to $w=79.2$ nm.
The effective confinement potential is plotted with the blue levels
(the lowest level corresponds to 10 meV, and the next are spaced by 40 meV).
The colorscale is given in meV.
(b) Harmonic oscillator potential (dots) for $\hbar\omega_0=2.9$ meV and the effective potential (solid line)
calculated along $y=-600$ nm line of panel (a).
(c) The thick curves shows the boundaries of the confinement region and the contour plot
shows the ground-state wave function bound at the junction of the ring to the leads.}
  \label{sys}
\end{figure}

\section{Theory}
We consider a quantum ring of a radius 155 nm with three symmetrically attached terminals - see Fig. \ref{sys}(a).
The simulations are based on the solution of the time-dependent Schr\"odinger equation
$i\hbar \frac{\partial \Psi}{\partial t}=H\Psi$,
for the Hamiltonian \begin{equation}H=\left(-i\hbar\nabla+e{\bf A}({\bf
r})\right)^2/{2m^*}\label{ham}\end{equation} in which we apply the Lorentz gauge ${\bf A}=(-By,0,0)$ and the GaAs effective mass
$m^*=0.067 m_0$. The problem is solved with a technique previously used in Refs. [\cite{time,epl,kali}] in which
the wave function is expanded in a basis
\begin{equation}
\Psi(x,y,t)=\sum_j c_j(t) f_j(x,y), \label{basis}
\end{equation}
of Gaussian functions $f_j$ localized around centers  $(X_j,Y_j)$
\begin{eqnarray}
f_j(x,y)&=&C\exp\left[-\frac{1}{2}m^*\omega_0\left((x-X_j)^2+(y-Y_j)^2\right)\right.  \nonumber \\
&&\left.+\frac{ieB}{2\hbar}(x-X_j)(y+Y_j)\right], \label{gf}
\end{eqnarray}
where $C$ is the normalization constant, $\omega_0$ determines the localization of the basis functions and
the imaginary term in the exponent introduces the magnetic translation phase shift that guarantees
the gauge invariance, i.e. the equivalence of all the applied centers in external magnetic field.

The applied choice of centers is shown by the dots in Fig. \ref{sys}(a).
The centers are spaced by 22 nm along the leads which is close enough to allow for a smooth electron flow along the channel
provided that the wave vector is lower than 0.15 / nm.
The electron wave function is confined in the direction perpendicular to the axis of the leads,
and the channel width can be estimated as $w=4 \sqrt{\hbar /m^* \omega_0}$.
The present modeling of the leads as a chain of functions (\ref{gf}) limits the simulation to the lowest subband.

The Hamiltonian (\ref{ham}) does not explicitly contain any confinement potential. In the present model the electron confinement results from the localization
of the Gaussian functions (\ref{gf}). Nevertheless, one can try to extract an effective confinement potential
present within the model by considering the Hamiltonian eigenstates obtained in basis (\ref{basis}). For that purpose we take the
 eigenequation $H\Psi_n=E_n\Psi_n$ and plug expansion (\ref{basis}) for the eigenstate $\Psi_n$. Subsequently, the weak form of the eigenequation
is obtained by its projection on the basis  elements (\ref{gf}), which produces the generalized eigenvalue problem ${\bf H} {\bf c}_n=E_n {\bf S} {\bf c}_n$,
where ${\bf H}$ and ${\bf S}$ are the Hamiltonian and overlap matrices with analytically integrable elements ${\bf H}_{lj}=\langle f_l|H| f_j\rangle$ and
${\bf S}_{lj}=\langle f_l| f_j\rangle$, respectively.
The effective potential in a point $(x,y)$ is then estimated by
\begin{equation}V(x,y)=\frac{\left[E_n -\left(-i\hbar\nabla+e{\bf A}\right)^2/2m^*\right]\Psi_n(x,y)}{\Psi_n(x,y)}.\end{equation}
The effective potential is plotted in Fig. \ref{sys}(a) for $\hbar\omega_0=2.9$ meV with the blue contour plot. Additionally a cross section of Fig. \ref{sys}(a)
calculated along the $y=-600$ nm line is plotted in Fig. \ref{sys}(b) with the solid line. The dots in Fig. \ref{sys}(b) show
the harmonic oscillator confinement potential $m^* \omega_0^2 x^2/2$. We can see that the effective confinement potential is consistent
with the nominal value of $\omega_0$  applied in the Gaussian basis (\ref{gf}). In Fig. \ref{sys}(a) we marked the channel region
in green. The green area in the figure was determined as
the one in which the sum of the Gaussian basis functions exceeds 10\% of its maximum value, which well agrees with the value of the channel width $w=79.2$ nm
obtained for $\hbar\omega_0=2.9$ meV.

The junctions of the wire to the ring allow for formation of bound electron states.
We find three nearly degenerate bound states -- a single bound state for a single junction --
in consistence with the known property of connections in the T-wires.\cite{bqd}
The binding energy is equal to  0.1 meV and the wave function of the lowest-energy bound state is plotted in Fig. \ref{sys}(c).
For the defect potential we use  $V_d(x,y)=W\exp\left[-((x-X_c)^2+(y-Y_c)^2)/R_d^2\right]$, where $(X_c,Y_c)$ are the coordinates
of the center of the defect, $R_d=28$ nm is its radius, and $W$ its height / depth.

As the  initial condition for our calculation we take a Gaussian wave function entirely localized in the
input lead (the one below the ring with axis $x=0$) localized in the direction perpendicular to the axis as the basis elements (\ref{gf})
but with a larger spread along the lead
\begin{eqnarray}
&&\Psi({\bf r},t=0)=f_j(x,y)\nonumber \\&& \times \exp\left(+\frac{1}{2}m^*\omega_0(y-Y_j)^2-\frac{\Delta k^2}{4}(y-Y_j)^2\right)  \nonumber \\
&&  \times \exp(iqy),
\label{ic}
\end{eqnarray}
where the last term in Eq. (\ref{ic})
is introduced to push the electron in the direction of the ring
with an average momentum $\hbar q$.
For the applied gauge the kinetic and canonical momentum in the $y$ direction are identical
and the $y$ component of the initial
probability density current integrated over the channel is equal to $\hbar q/m^*$.
The Fourier transform of the initial condition along
the axis of the lead is
$\tilde{\Psi}(k)=\sqrt{\pi/2}\Delta k\exp(-{(q-k)^2}/{\Delta k^2})$, and $\Delta k$ is interpreted as the dispersion of the packet in the wave vector space.
The initial condition (\ref{ic}) is projected onto the basis ($\ref{basis}$), and the rest of calculation amounts in determining the coefficients $c_j(t)$
in subsequent moments in time. We use the matrix version of the Askar-Cakmak scheme \cite{time,askar}
in form of a system of linear equations for ${\bf c}(t+dt)$,
\begin{equation} {\bf S}{\bf c}(t+dt)={\bf S}c(t-dt)-\frac{2idt}{\hbar}{\bf H}{\bf c}(t).\label{soe}\end{equation}
We use $dt=0.01$ ps. Reduction of the time step below this value does not change the results.

In the present approach the transfer probabilities $T_l$ and $T_r$ are determined by the parts of the packet
which are transferred to the leads before the end of the simulation. The simulation is terminated
when the electron packet completely leaves the ring. We consider the ring as empty when
it does not contain more than $0.001\%$ of the electron charge.
Generally, in the time-dependent calculations the transferred and back-scattered wave packets
return to the ring after reflection from the ends of the channels unless absorbing \cite{pml}
or transparent \cite{tns} boundary conditions are used. Application of open boundary conditions
is crucial for approaches using finite difference techniques.\cite{sgs,chaves}
The present work could be performed without any open
boundary conditions since in the present approach one can apply leads of a length which is in practice arbitrarily large.
A LU decomposition of the  ${\bf S}$ matrix for the system of equations (\ref{soe}) is performed before the time stepping.
With the decomposed overlap matrix the numerical cost of each time step scales linearly with
the number of centers. For the present calculation we use in total 5000 Gaussian functions (\ref{gf}) with the leads as long as 30 $\mu$m each.

In strictly one-dimensional modeling of quantum rings with the scattering matrix formalism \cite{buttikernowy} the
ring and the leads are treated as separate objects with the coupling strength described by an appropriate parameter.
The coupling strength is responsible for the time spent by the electron within the ring and the sharpness
of the transfer probability extrema in function of the wave vector. In the present two-dimensional model the ring  and the
leads are modeled as a single object and the ring is essentially open. Nevertheless the
junctions of the ring to the leads act like small scattering cavities.\cite{sols}
In this paper the junctions are modeled
as right-angle connections.  Smooth junctions -- binding more than a single electron state -- were recently studied in Ref. [\onlinecite{chaves}].
The type of the junction affects the wave vector dependent transfer probabilities, but their magnetic-field behavior remains
qualitatively unchanged.

\section{Results and discussion}

We first briefly present the results obtained for an ideally symmetric configuration (section III.A) to set
the reference point for the discussion of the transport for a defect present in one of the arms of the ring (section III.B).
\subsection{Clean ring}
\begin{figure}[ht!]
\epsfysize=65mm \epsfbox[14 160 577 711]{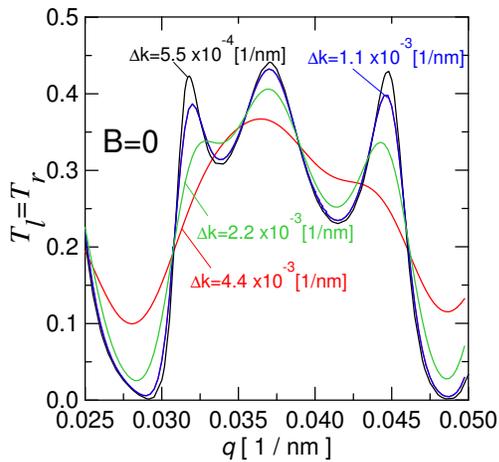} \\
\caption{Electron transfer probability to the left $T_l$ and right $T_r$ leads for $B=0$
in function of the average wave vector $q$ for a number of $\Delta k$ values for $w=79.2$ nm.}
  \label{todqwk}
\end{figure}

In order to set the wave vector dispersion parameter $\Delta k$ close enough to the monoenergetic limit
we studied the transmission probability through an ideally symmetric ring for $B=0$ -- see Fig. \ref{todqwk}.
In the small $\Delta k$ limit we notice that
the variation of $T(q)$ becomes more pronounced and the dependence distinctly saturates.
The main features
of the saturated $T(q)$ dependence are well resolved already for the $\Delta k=1.1\times 10^{-3}$ /nm, the value which is used in the rest of the paper.
Below   -- unless explicitly stated otherwise -- we also assume $q=0.037$ /nm as the average
wave vector -- the value which corresponds to a maximum of $T(q)$ and $w=79.2$ nm.
The kinetic energy of progressive motion
equals $\hbar^2 q^2/2m^*=0.77$ meV, and the dispersion of the wave vector assumed covers the energy
window $(\hbar^2 (q-\Delta k)^2/2m^*,\hbar^2 (q+\Delta k)^2/2m^*)=(0.73,0.82)$  meV.
The second subband is by $\hbar \omega_0=2.9$ meV above the lowest one --
far above the electron kinetic energy -- which is consistent with the neglect of the scattering
to higher subbands assumed in the present approach.

\begin{figure}[ht!]
\epsfysize=65mm \epsfbox[24 170 548 706]{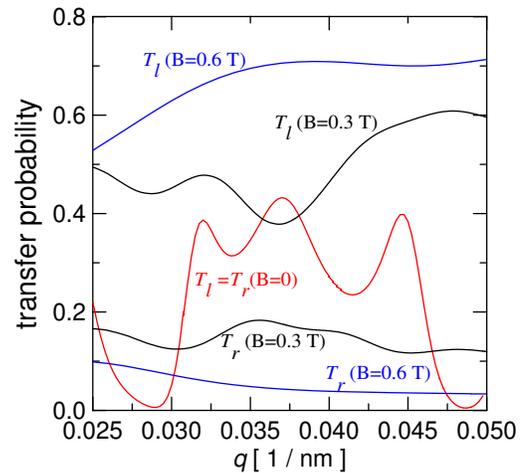} \\
\caption{Electron transfer probability to the left $T_l$ and right $T_r$ lead for $B=0,0.3$ and 0.6 T
in function of the average wave vector $q$ for $\Delta k=1.1\times 10^{-3}$ /nm.}
  \label{todqwb}
\end{figure}

\begin{figure}[ht!]
\begin{tabular}{ll}
a) & \epsfysize=60mm \epsfbox[24 92 592 621]{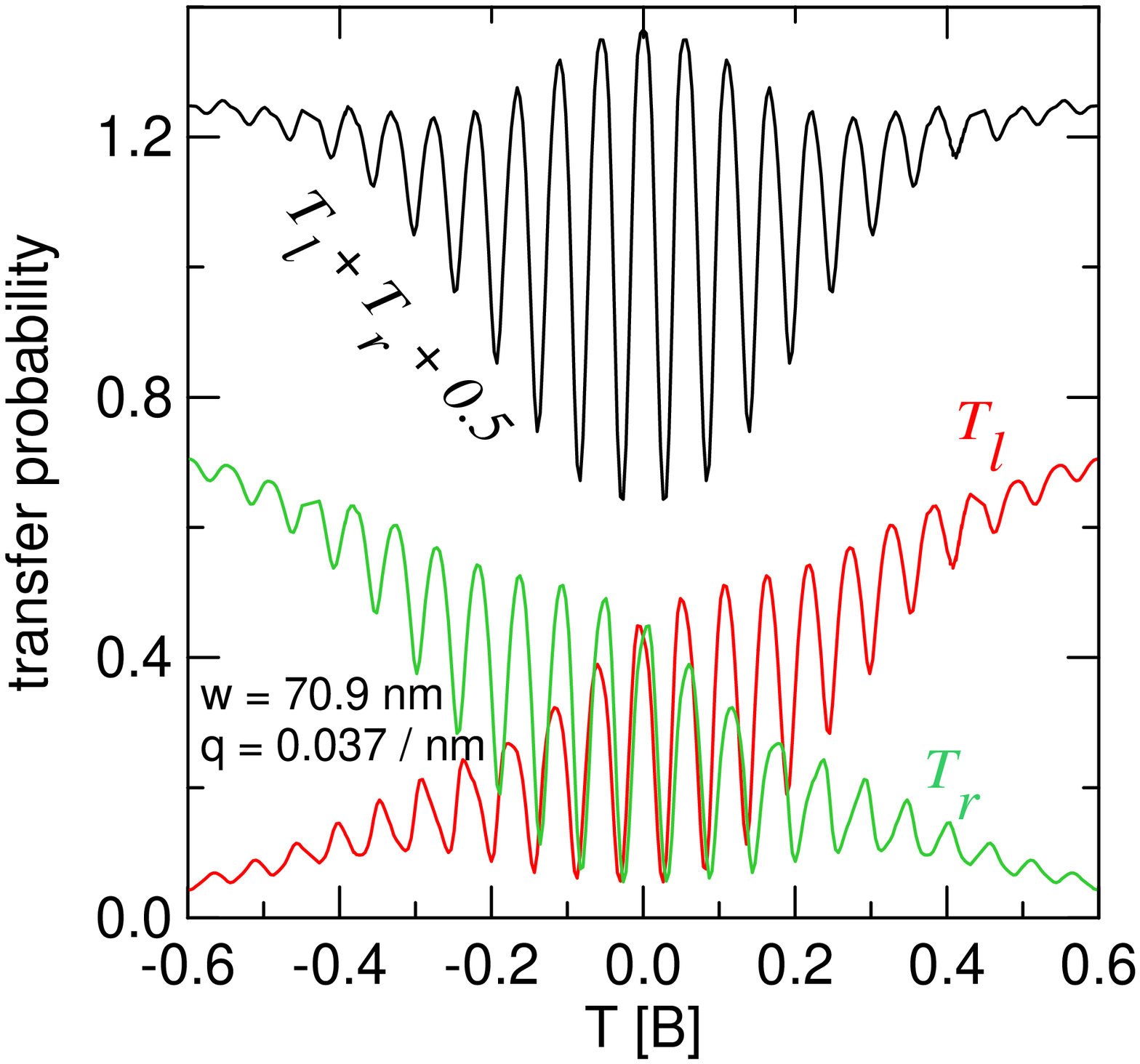} \\
b) & \epsfysize=60mm \epsfbox[24 92 592 621]{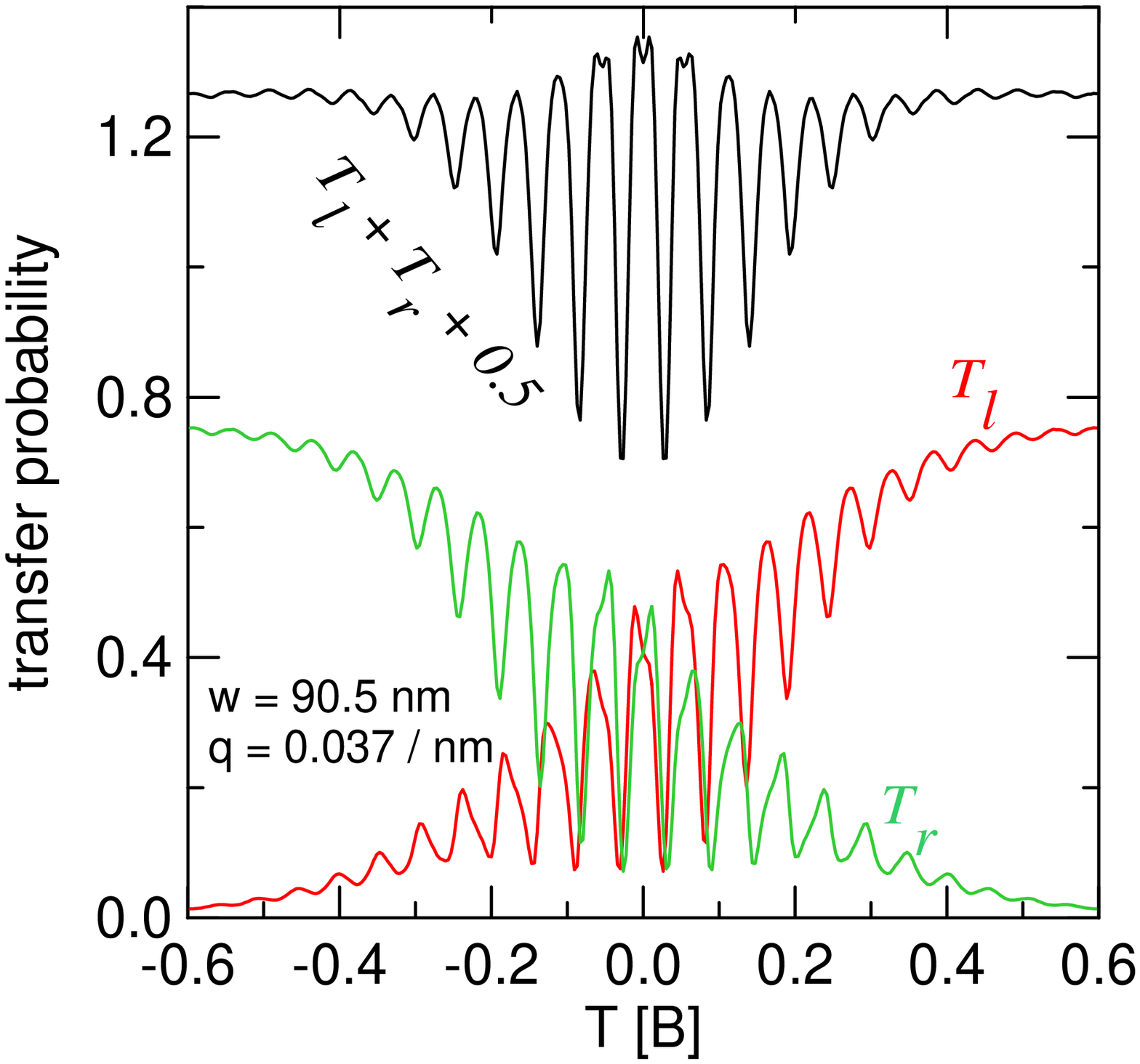} \\
c) & \epsfysize=60mm \epsfbox[24 92 592 621]{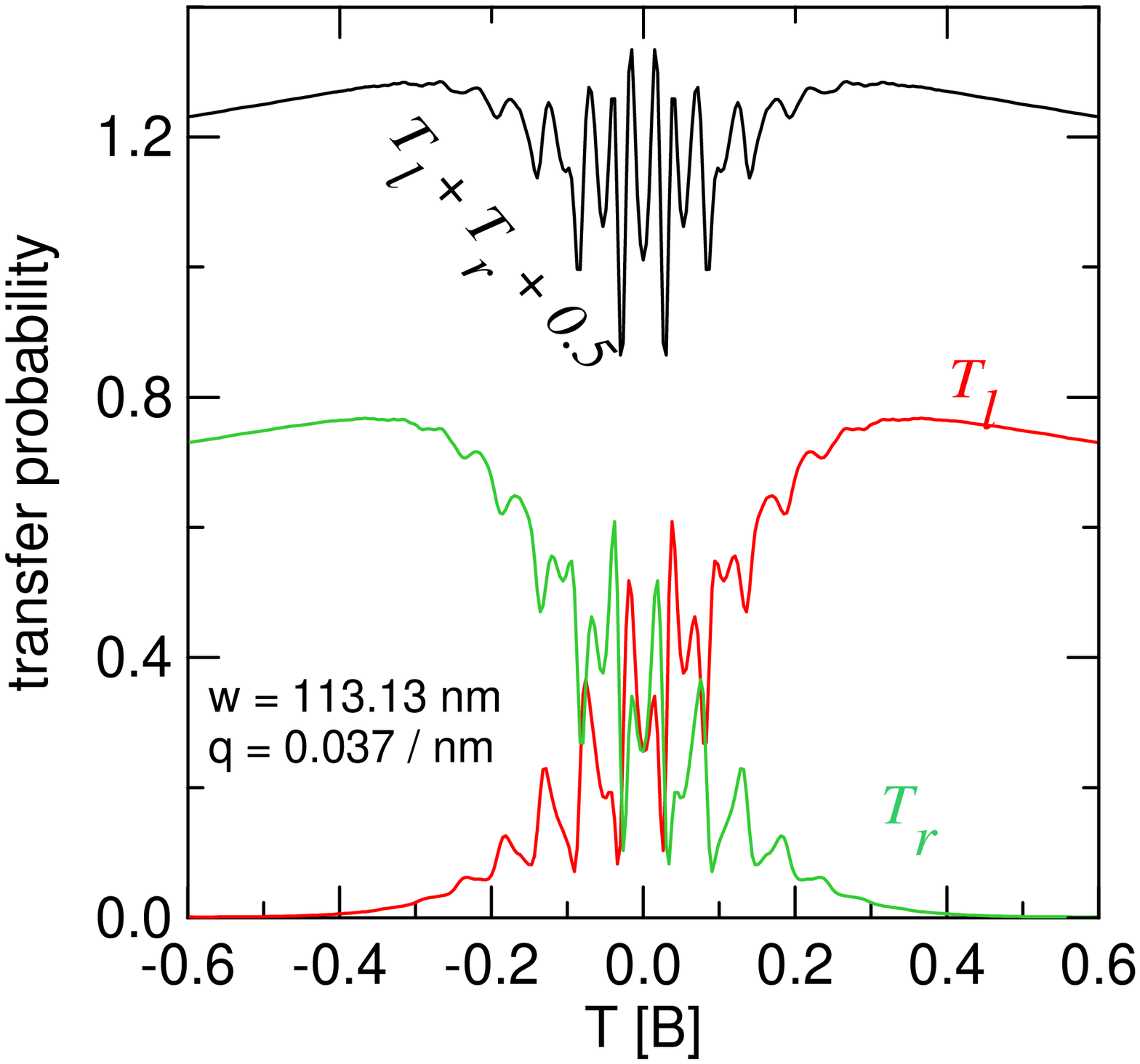} \\
\end{tabular}
\caption{Electron transfer probability to the left $T_l$ and right $T_r$ leads
as well as their sum (shifted up for clarity by 0.5) for pure ring and channels of width $w=79.2$ nm (a)
$w=90.5$ nm (b) and $w=113.13$ nm (c).  }
  \label{wBws}
\end{figure}

The effect of the magnetic field on the wave-vector resolved transfer probability through the symmetric ring is illustrated
in Fig. \ref{todqwb}. We notice that the external magnetic field introduces
asymmetry of the transfer to the left and right output leads. For $B>0$, $T_l$ increases
at the expense of the $T_r$, which is a direct consequence of the magnetic forces which
preferentially inject the electron to the left arm of the ring and then eject it to the left output lead.
For higher magnetic fields $T_l$ and $T_r$ become less strongly dependent on $q$.

The dependence of the transfer probability on $B$ is shown in Fig. \ref{wBws}(a).
The transmission probabilities  $T_l$, $T_r$ as well as their sum $T$  undergo oscillations that are due to the Aharonov-Bohm effect.
The period of the oscillations of $T$ is equal to 0.055 T, in agreement
with the nominal value of the magnetic field, which corresponds to the flux quantum $\Phi_0=e/h$ for the ring of radius 155 nm.
The distinct decrease of the oscillations amplitude for larger $B$ is another consequence of the magnetic forces.
For higher fields most of the electron packet is injected into one of the arms of the ring, which subsequently introduces an
imbalance in the parts of the packet that meet and interfere near the output leads.
In consequence the Aharonov-Bohm interference is less pronounced.

In order to demonstrate the effect of the channel width we presented in Fig. \ref{wBws}(b) and (c)
the transfer probabilities for wider channels, namely for  $w=90.5$ nm (b)
and $w=113.13$ nm (c) (in Fig. \ref{wBws} and everywhere else in this paper we assume $w=79.2$ nm).
The strength of the magnetic deflection of the electron trajectories depends on the ratio of the Larmor radius to the channel width
-- for wider channels there is more space for the magnetic deflection, and consequently
the Aharonov-Bohm oscillations vanish faster for larger $w$.

In Fig. \ref{todqwb} we also notice, that even for the symmetric ring the extrema of $T_l$ and $T_r$ are shifted off $B=0$ (in other words $\frac{\partial T_{l/r}}{\partial B}|_{B=0}\neq 0$)
only the sum of the transmission probabilities $T(B)=T_l(B)+T_r(B)$ is an even
function of $B$. For the symmetric ring one has \begin{equation} T_l(B)=T_r(-B),\label{ts}\end{equation} which suffices to explain the $T(B)=T(-B)$ symmetry
observed in Fig. \ref{todqwb}. However, for asymmetric rings relation  (\ref{ts}) no longer holds, although the transfer probability remains symmetric in $B$ -- see below.

\subsection{Ring with a defect}

\begin{figure}[ht!]
\epsfxsize=50mm \epsfbox[33 140 562 670]{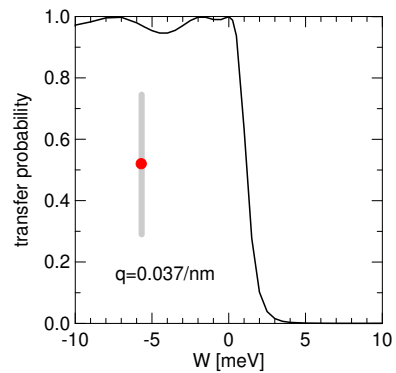}
\caption{Electron transfer probability for a straight channel containing
a Gaussian defect of radius $R_d=28$ nm and height/depth $W$ (negative values of $W$ correspond to potential well), for $B=0$.}\label{str}
\end{figure}

Before inserting the Gaussian defect to the ring we first study its scattering properties for a straight channel. Fig. \ref{str} shows the transmission probability as  function of the height
of the defect $W$ for the average wave vector $q=0.037$ / nm. For the attractive defect ($W<0$) the transmission probability is close to 1, and the defect acts like a phase shifter. For $W>0$ the defect is a more effective scatterer
with transmission probability as low as 0.08\% for $W=5$ meV.

\begin{figure*}[ht!]
\epsfxsize=120mm \epsfbox[12 478 806 806]{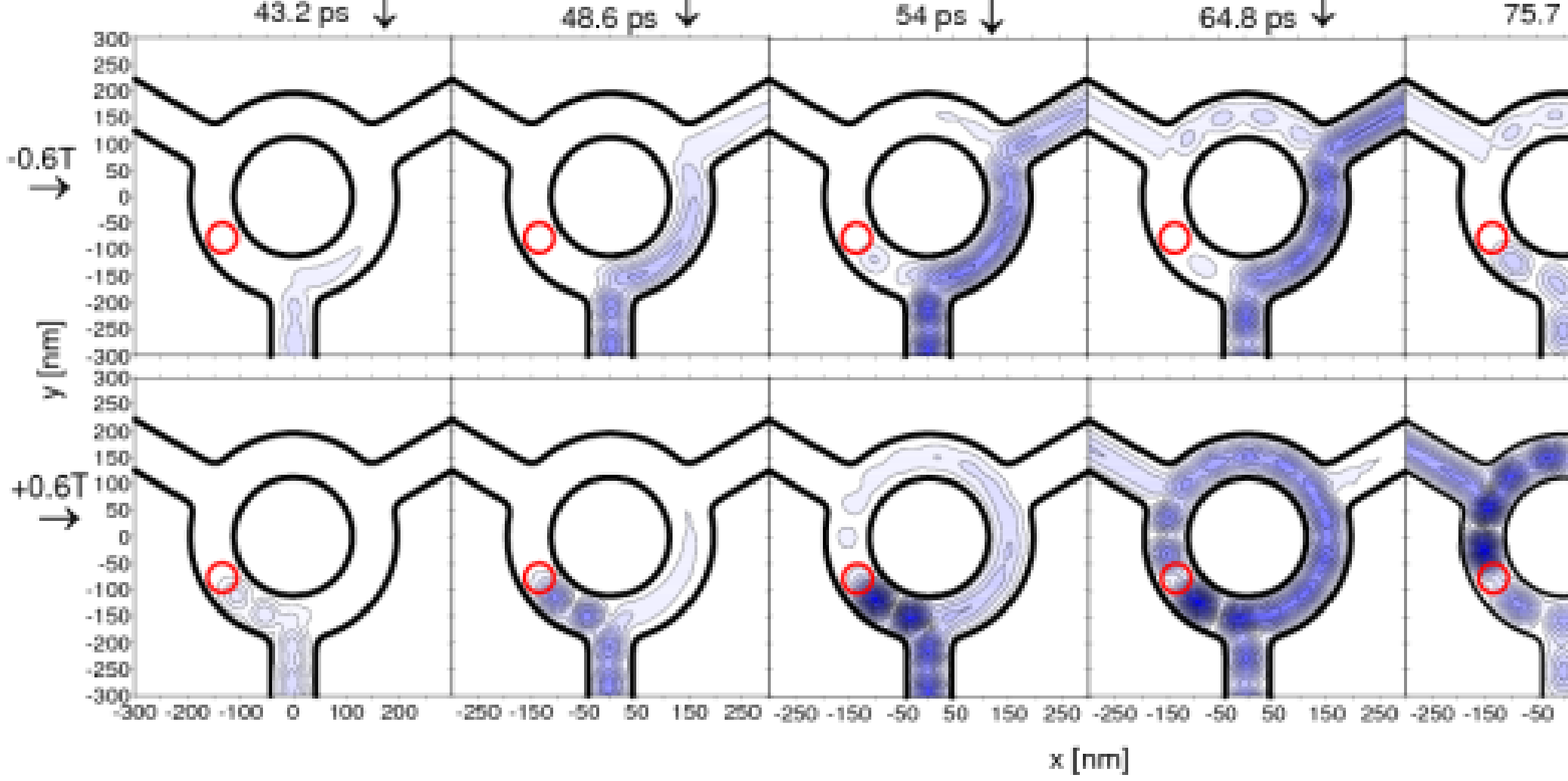}
\caption{Amplitude of the wave function calculated for $B=-0.6$ T (the upper row of plots) and $B=0.6$ T (the lower row)
at subsequent moments in time (given on top of the figure).
The circle is centered at the position of the repulsive defect ($W=3$ meV).\label{czas}}
\end{figure*}

\begin{figure}[ht!]
\begin{tabular}{ll}
\epsfxsize=30mm \epsfbox[110 324 562 755]{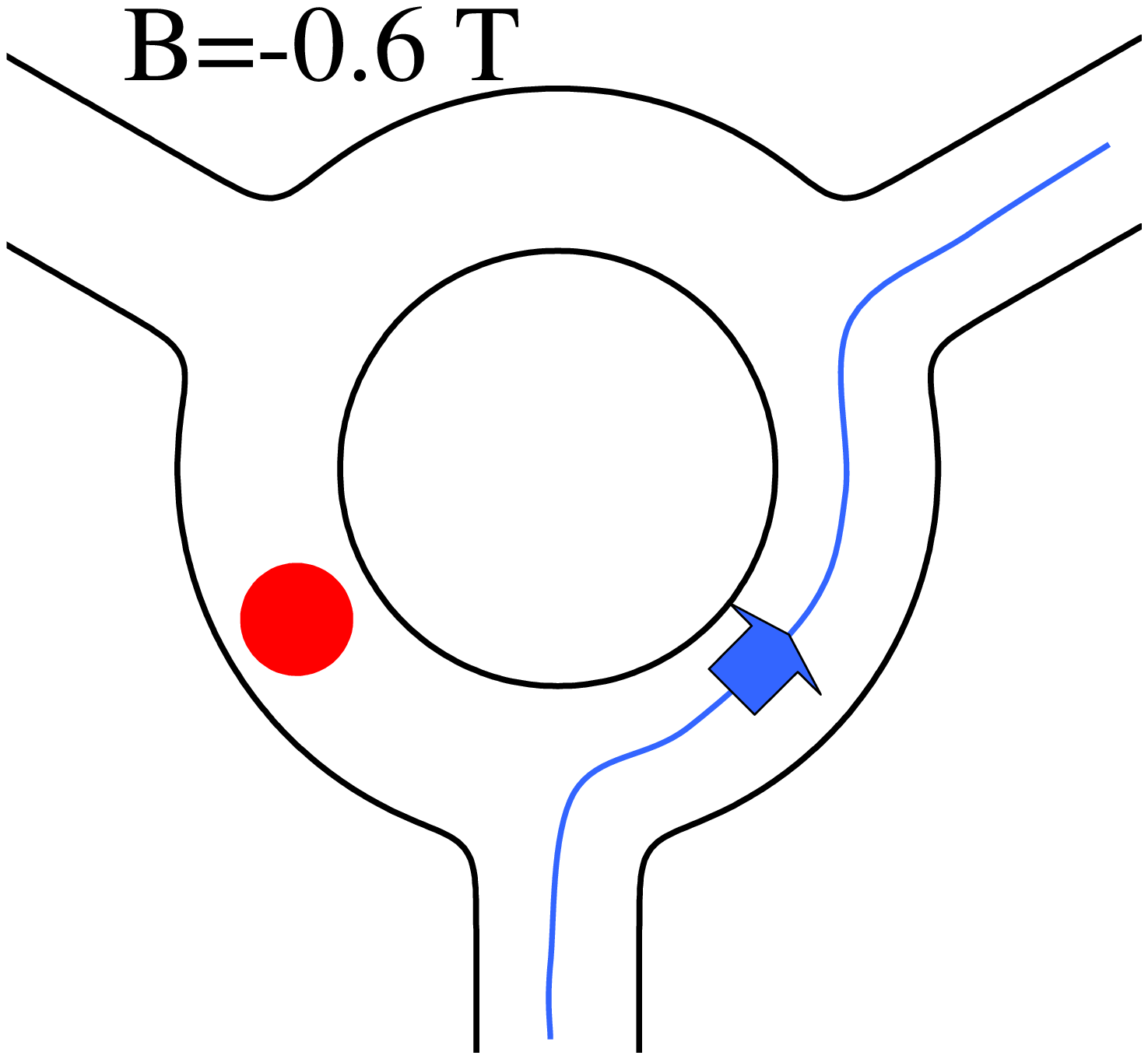} & \epsfxsize=30mm \epsfbox[110 324 562 755]{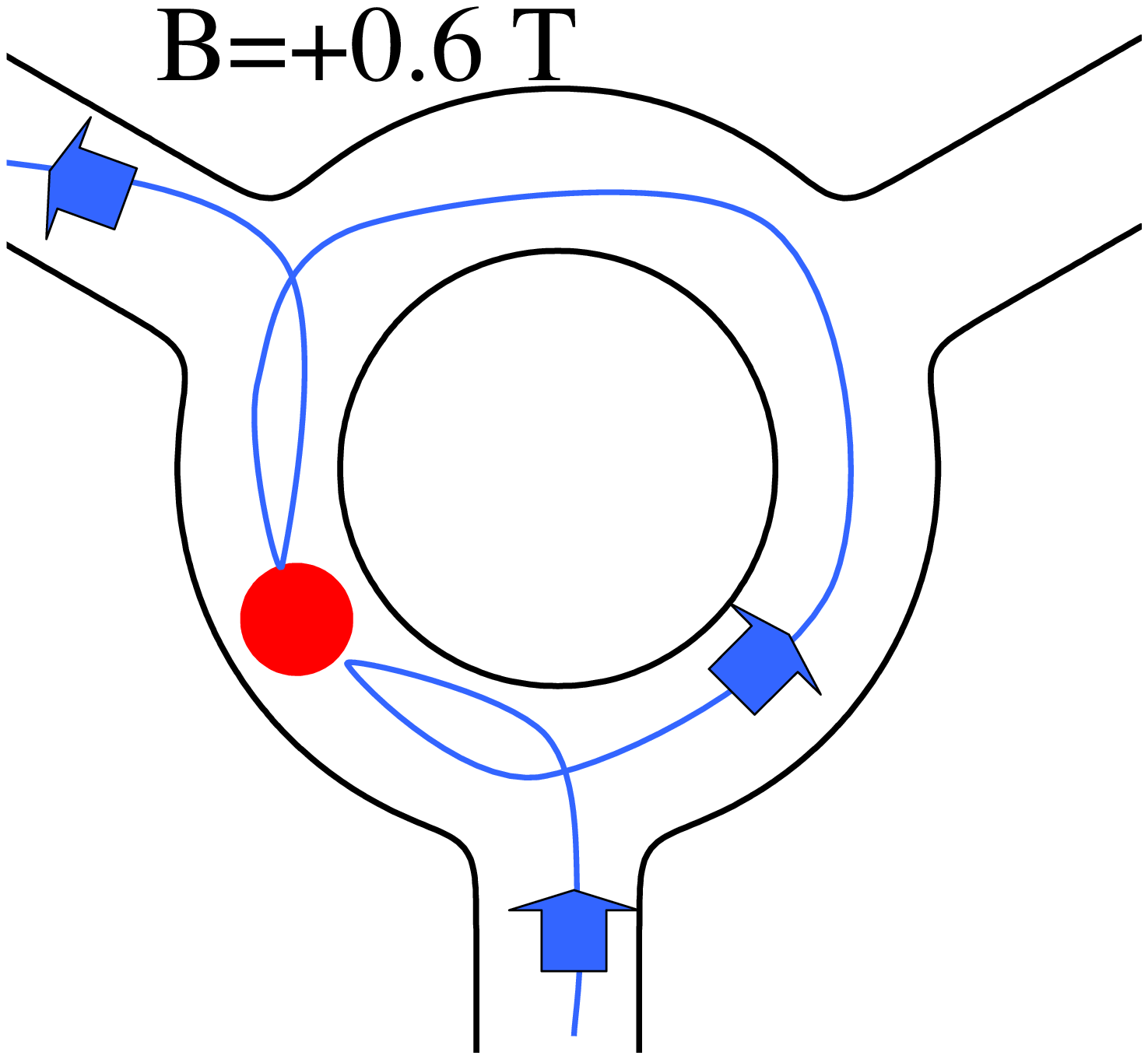} \\
a) & b) \\
\end{tabular}
\caption{Schematic drawing of the dominant trajectory in the electron transport presented in Fig. \ref{czas} \label{schemat}
for $B=-0.6$ T (a) and $B=0.6$ T (b).}
\end{figure}

\begin{figure}[ht!]
\epsfxsize=60mm \epsfbox[33 65 562 605]{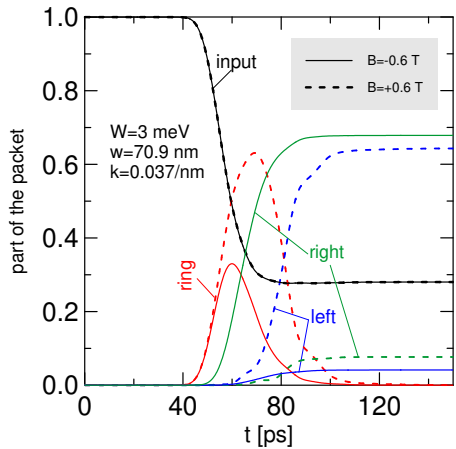}
\caption{Part of the wave packet inside (red curves) the ring, in the input lead (black curves),
in the left (blue curves) and right (green curves) output leads. Results for $B=0.6$ T are plotted with dashed
curves and for $B=-0.6$ T with solid ones. The data correspond to the wave function snapshots presented in Fig. \ref{czas}. \label{mielenie}}
\end{figure}

Next, we place a defect of height $W=3$ meV in the left arm of the ring between the input and the left output lead  -- see the
red circle in Fig. \ref{czas}, which also shows the amplitude of the electron wave function for $B=\pm 0.6$ T for several
moments in time. For the negative $B$ most of the electron wave function is injected to the right arm and
next to the right output lead, as for a pure quantum ring. The dominant trajectory for this field is drawn schematically
in Fig. \ref{schemat}(a). More complex is the transport for the positive magnetic field [the lower panel of Fig. \ref{czas} and Fig. \ref{schemat}(b)]:
the electron is first injected into the left arm, then it is nearly completely reflected by the defect. The electron velocity
is inverted and the Lorentz force -- still tending to deflect the trajectory to the left -- keeps the electron within the ring as it passes
near the input lead and then both output leads. The electron is subsequently reflected again from the defect, this time from its other side.
Only after the second scattering event the magnetic force pushes the electron to the left output ring.
In consequence most of the probability density goes to the left output lead just like for the pure ring.
In consistence with the schematic drawing of the dominant trajectory of Fig. \ref{schemat}(b) in Fig. \ref{czas} for $B=0.6$ T we first
observe an increased probability amplitude below the defect (48.6, 54, 64.8 ps) and then above it (75.7 and 86.5 ps).

Fig. \ref{mielenie} shows the parts of the electron packet within the ring and in the leads calculated for the simulation presented
in Fig. \ref{czas}. For the positive magnetic field the electron trajectory has a larger length  [Fig. \ref{schemat}(b)] and the electron packet
stays longer within the ring.  In the large $t$ limit we see in Fig. \ref{mielenie} that relation
(\ref{ts}) no longer holds. However, the part of the electron packet in the input lead (the incoming and backscattered
parts of the packet), is for any $t$ exactly the same for both magnetic field orientations. The independence of the backscattering
of the magnetic field orientation can be understood as due to the fact
that the backscattered trajectories are identical for $\pm B$.\cite{kali}
As a result of this invariance, the relation $T(B)=T(-B)$ holds in spite of the very different kinetics of the electron transfer
through the asymmetric ring for $B=\pm 0.6$ T.

\begin{figure*}[ht!]
\begin{tabular}{llll}
a) & b) &c) & d)\\
\epsfysize=37mm \epsfbox[24 109 540 610]{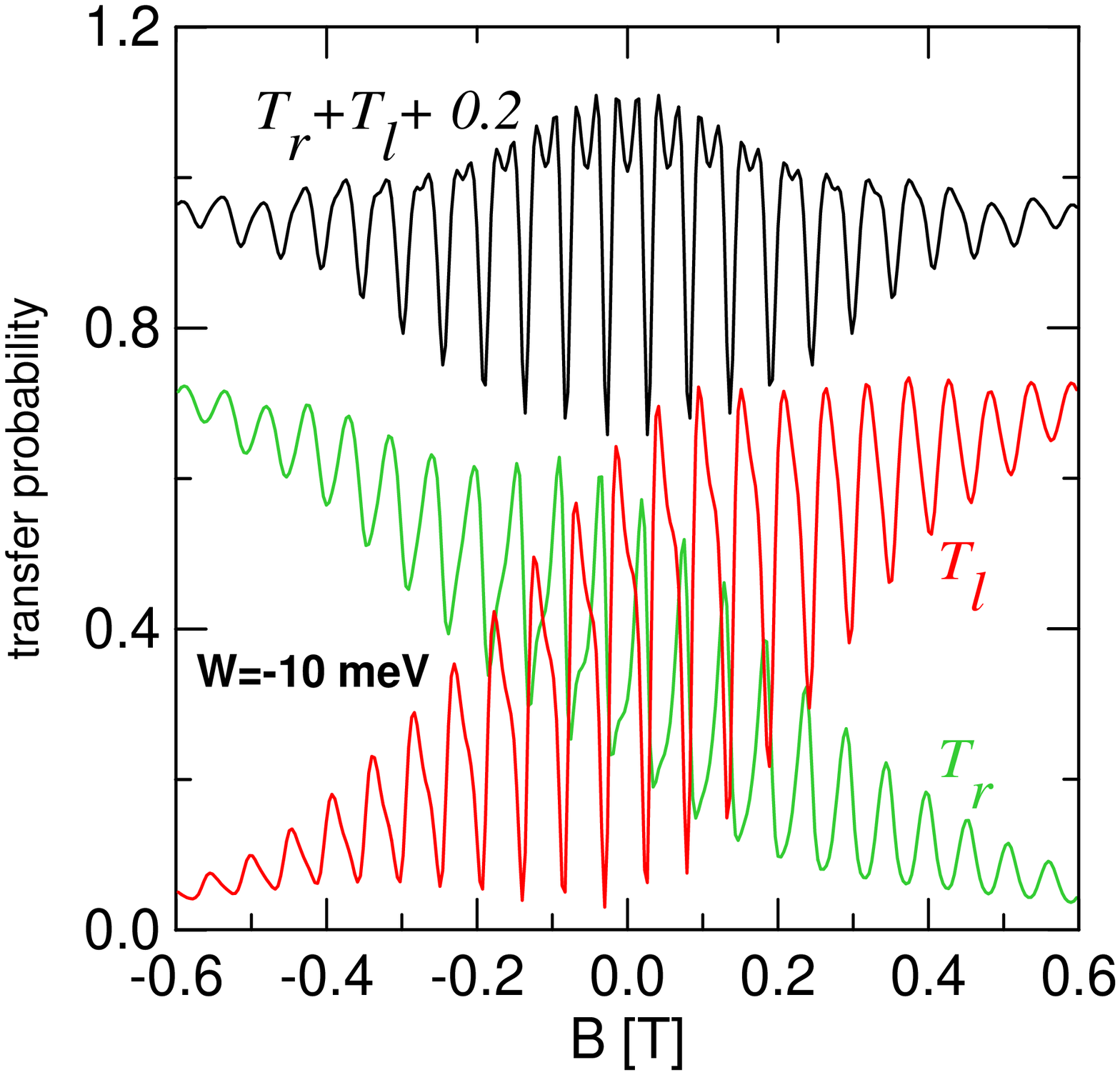} &
\epsfysize=37mm \epsfbox[24 109 540 610]{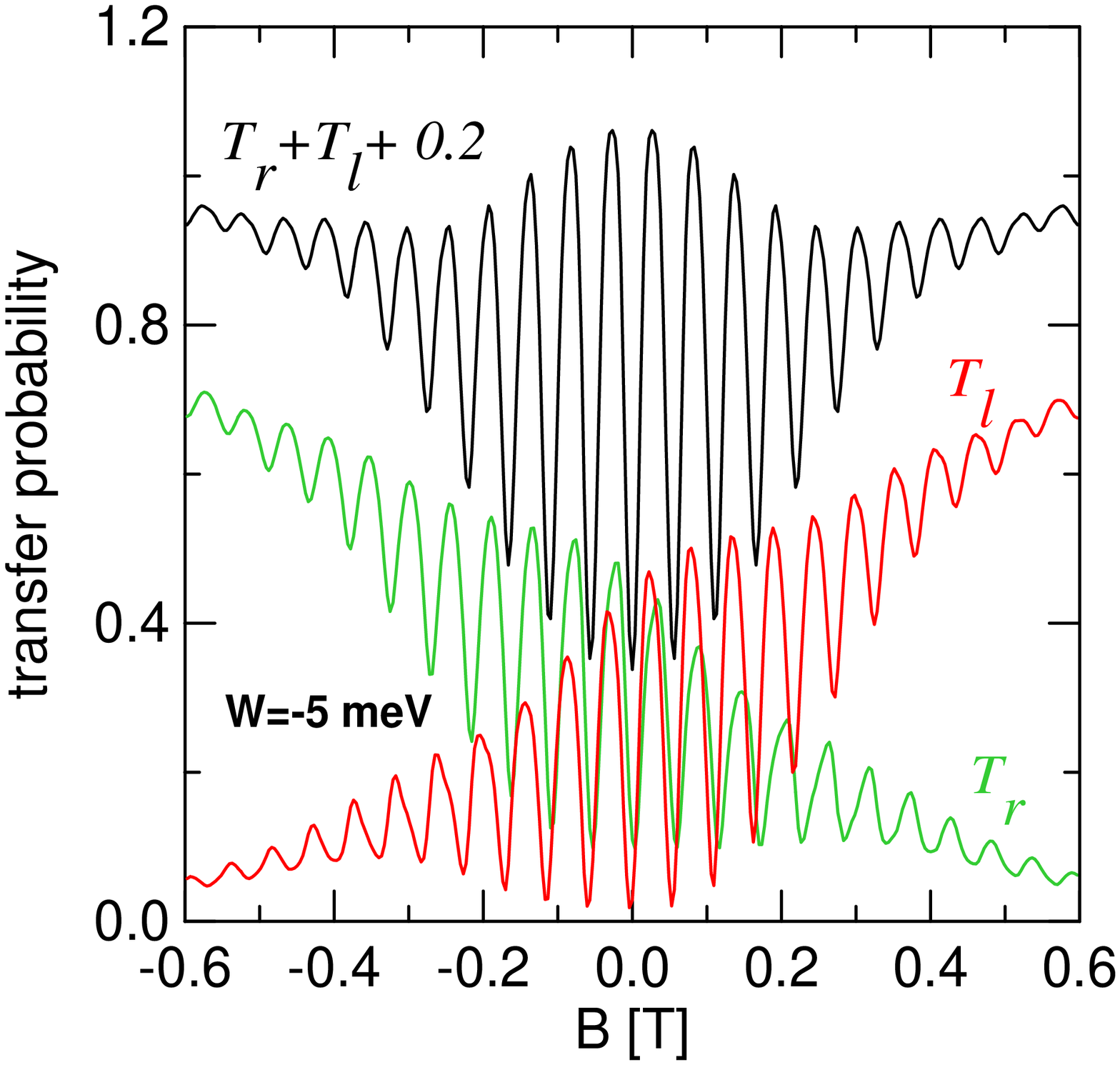} &
\epsfysize=37mm \epsfbox[24 109 540 610]{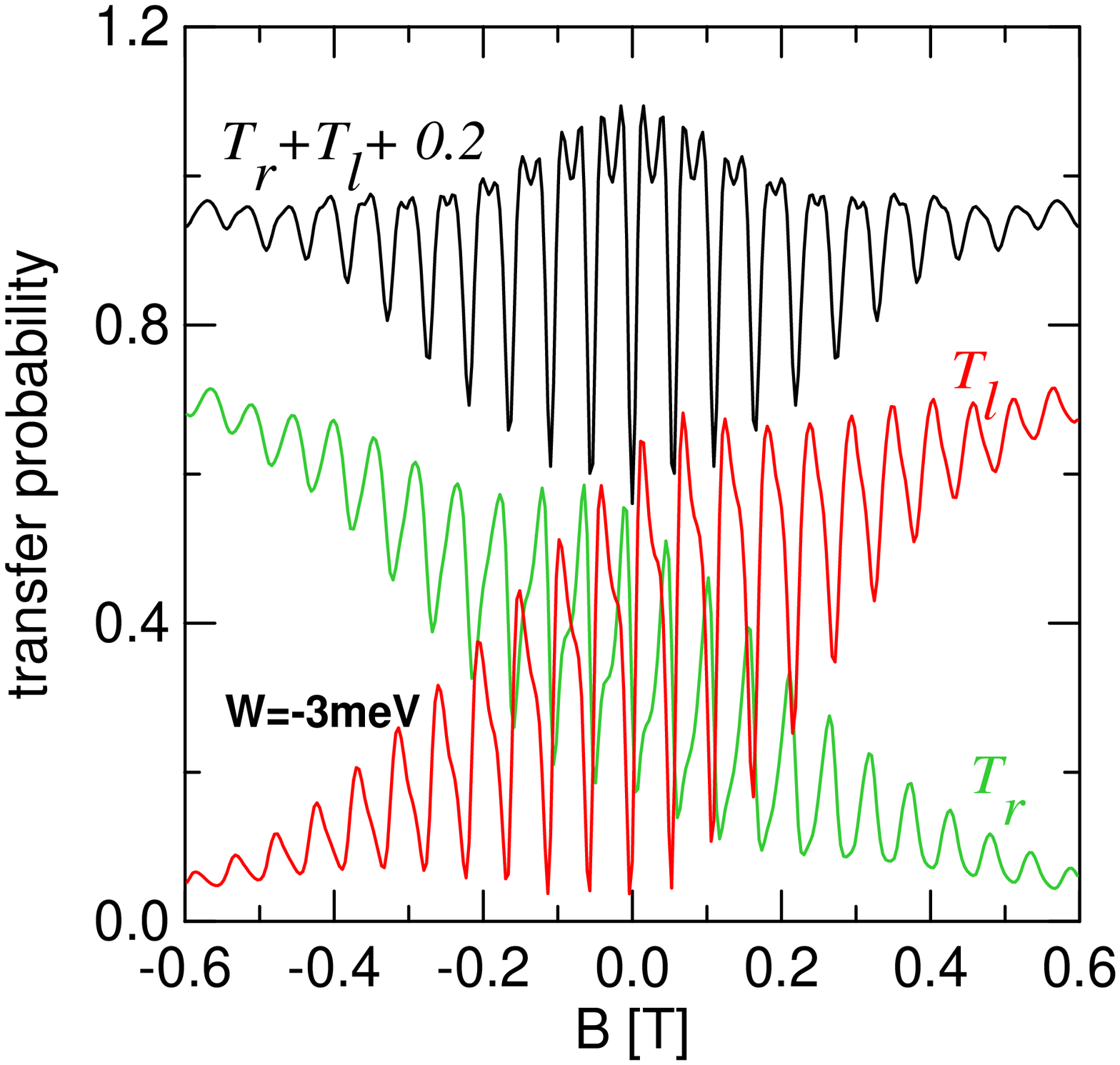} &
\epsfysize=37mm \epsfbox[24 109 540 610]{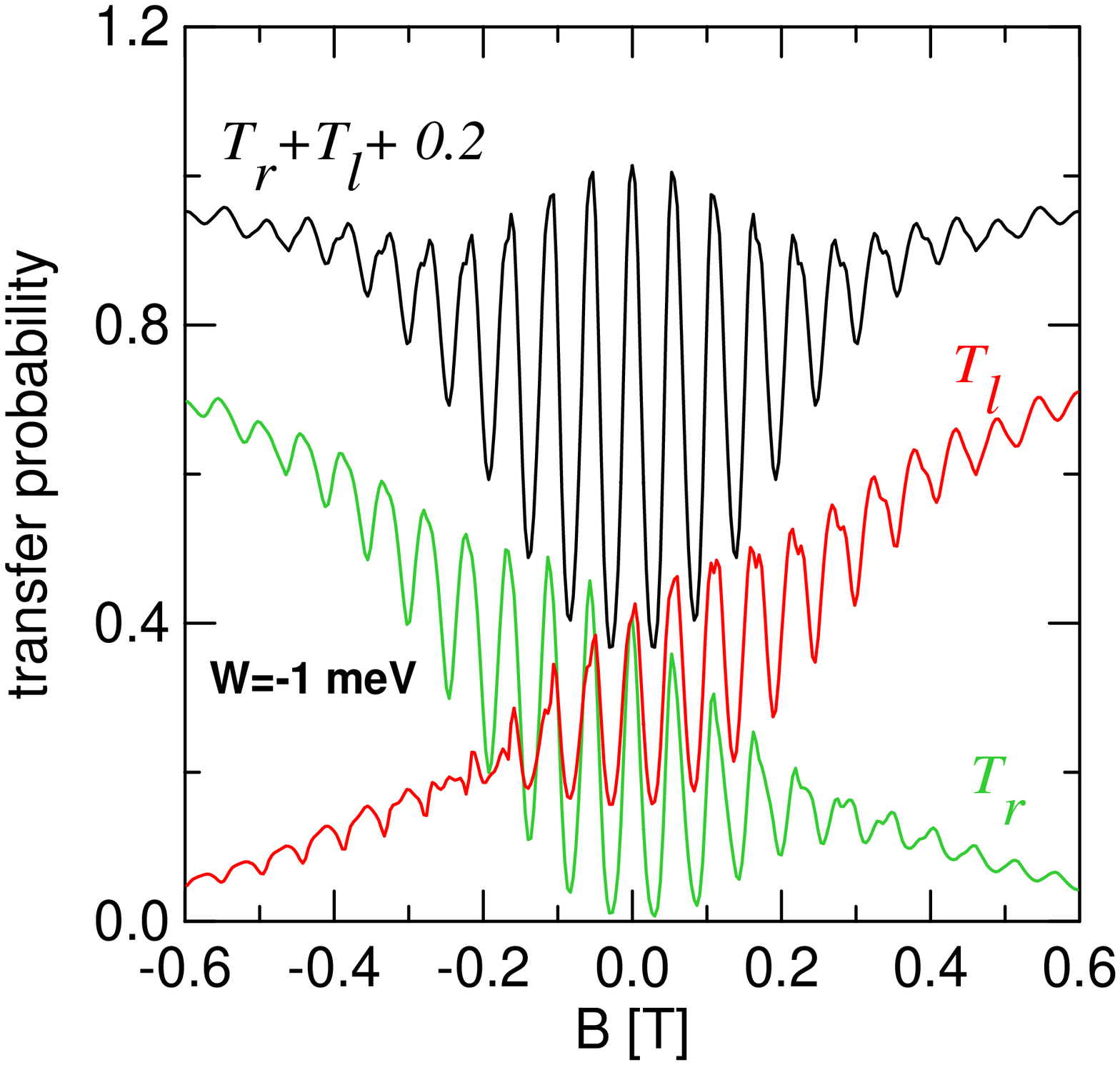} \\
e) & f) &g) & h)\\
\epsfysize=37mm \epsfbox[24 109 540 610]{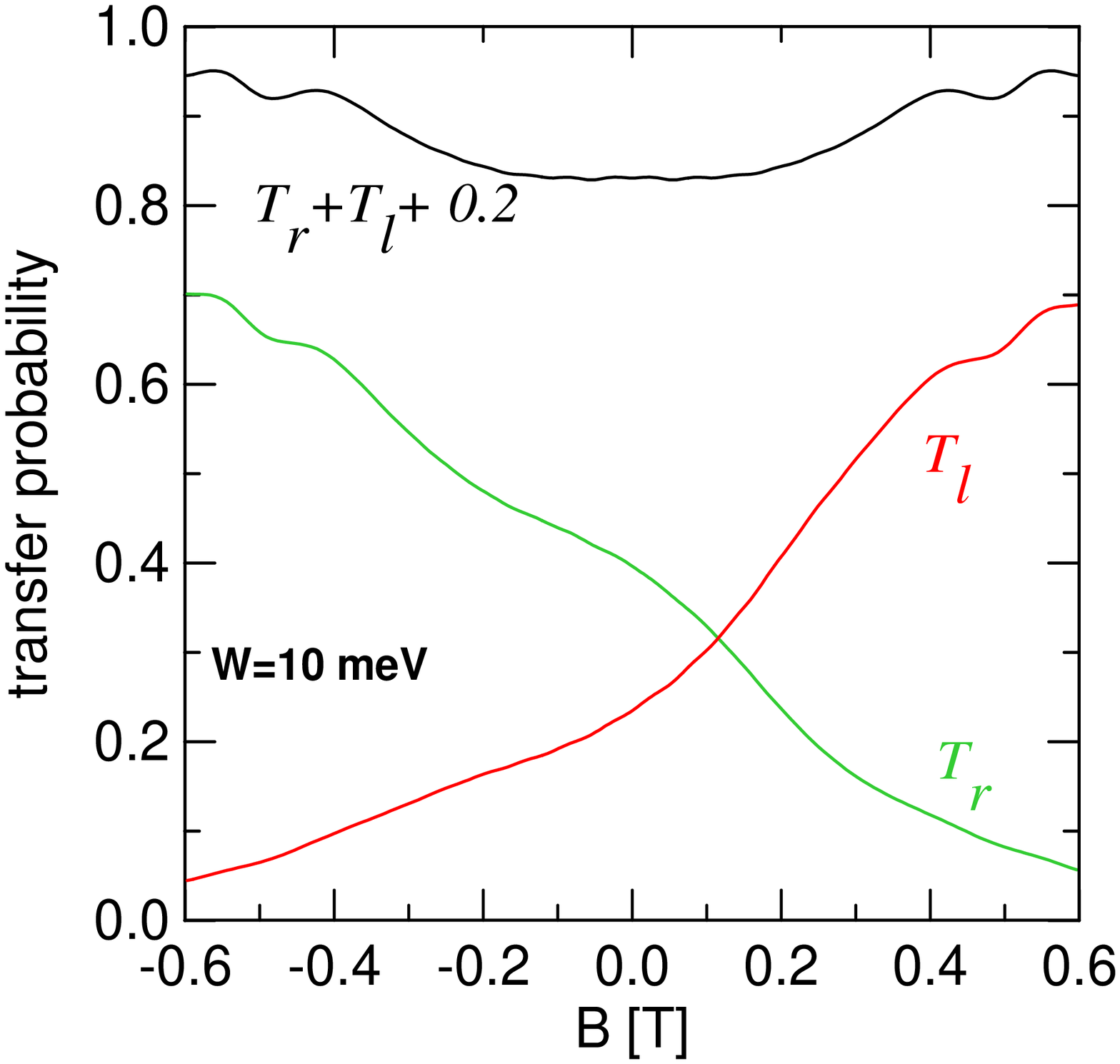} &
\epsfysize=37mm \epsfbox[24 109 540 610]{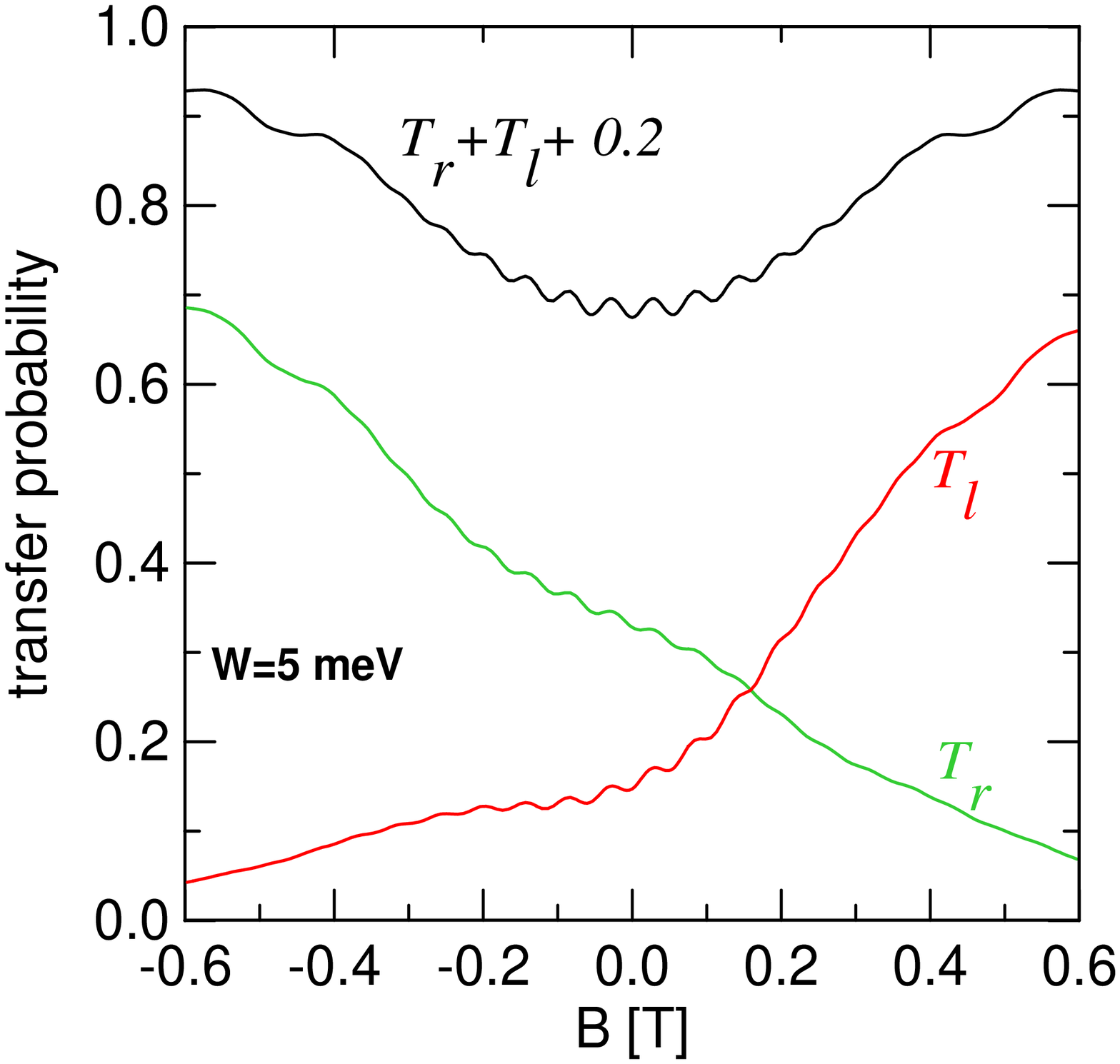}&
\epsfysize=37mm \epsfbox[24 109 540 610]{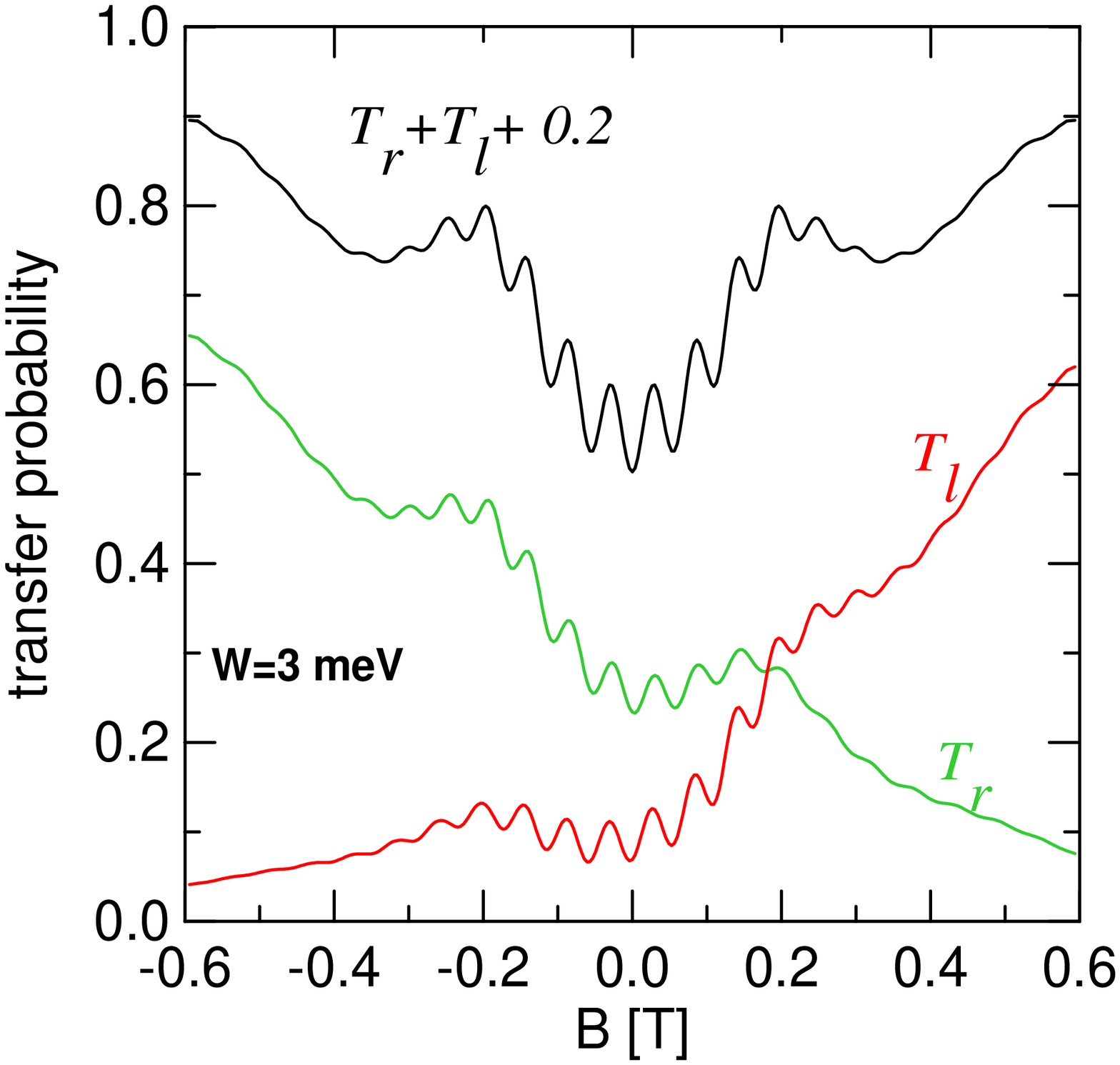}&
\epsfysize=37mm \epsfbox[24 109 540 610]{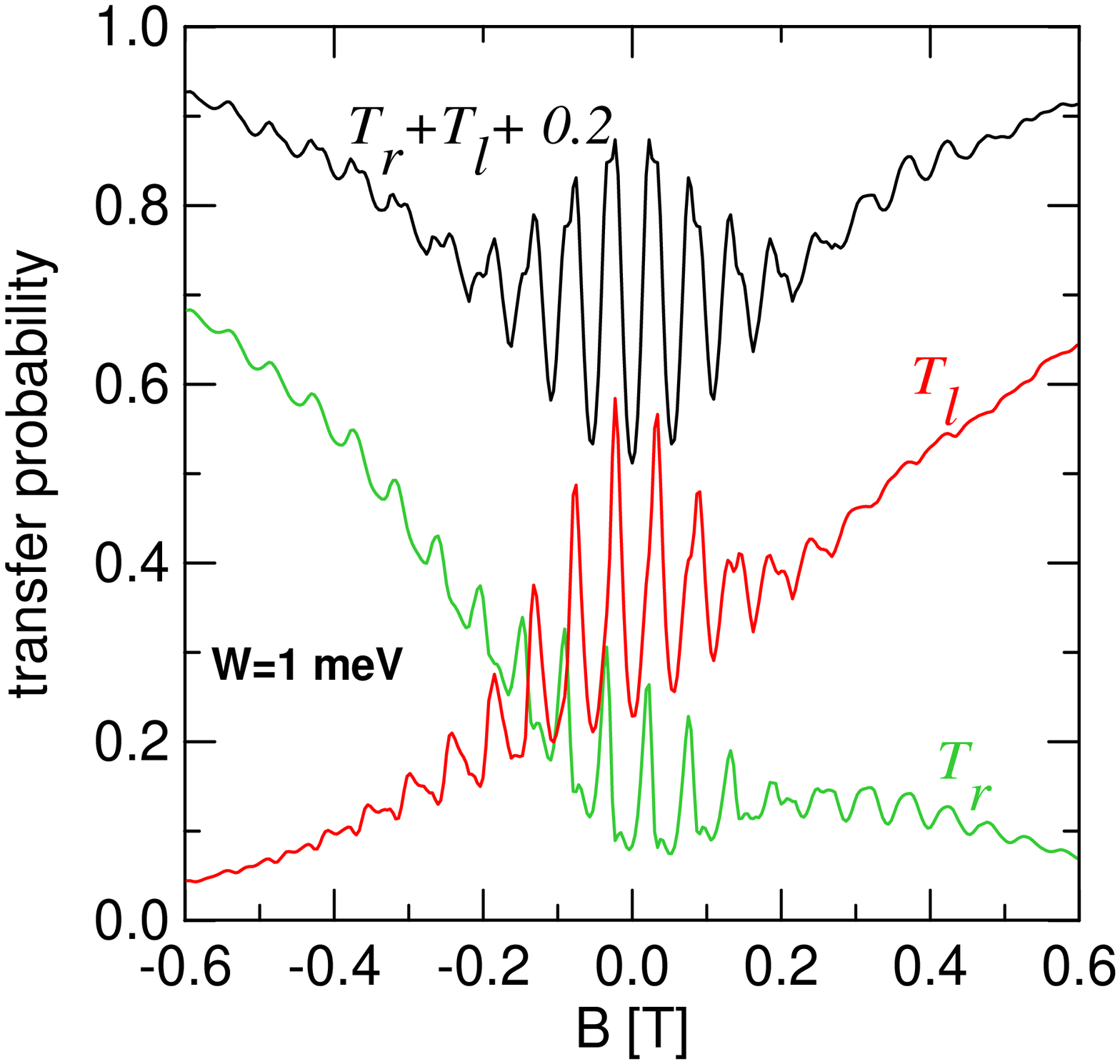}  \\
\end{tabular}
\caption{Electron transfer probability to the left $T_l$ and right $T_r$ leads
and their sum (shifted up for clarity by 0.2) for
the defect height $W=-10$ (a), -5 (b), -3 (c), -1 (d), 10 (e), 5(f), 3(g) and 1 meV (h).}
  \label{wv}
\end{figure*}

The oscillations of the transfer probabilities in function of $B$ are plotted in Fig. \ref{wv} for attractive (a-d) and repulsive  (e-h)
defects of different height / depth. The results can be compared with the ones given for the pure ring in Fig. \ref{wBws}(a),
and the parameters considered in Figs. \ref{czas}, \ref{schemat} and \ref{mielenie} are applied in Fig. \ref{wv}(g).
The presence of the attractive defect, which (only) shifts the phase of the part of the wave function passing through the left arm,
changes the local maximum of $T$ for $B=0$ [Fig. \ref{wBws}(a) and \ref{wv}(d)] into a local minimum [Fig. \ref{wv}(b) and (c)].
However, the average values of $T_l$ and $T_r$ within the range of small magnetic fields  $B \in [-0.2$ T, $0.2$ T$]$ remain very similar.
On the contrary, the repulsive defect [Fig. \ref{wBws}(e-h)] leads to a pronounced difference in $T_l$ and $T_r$ values for $B\simeq 0$. Moreover, since
the circulation of the electron around the ring is stopped by the repulsive defect (see Fig. \ref{czas}),
its presence drastically reduces the amplitude of the Aharonov-Bohm oscillations.

The results of Fig. \ref{wv}(b) are in a good agreement with the experimental results for conductance given in Fig. 1 of  Ref. [\onlinecite{wlochy}]
within the range of the magnetic fields presented therein: i) the amplitude of the oscillation is small, ii) for low magnetic fields $T_r$
distinctly exceeds $T_l$, iii)  $T_l$ dominates for $B>0$ and $T_r$ for $B<0$
iv) the oscillations of the transfer probabilities vanish at higher magnetic field, v) the overall transfer probability $T$
stays symmetric in $B$, vi) the envelope of the $T$ oscillations possesses a pronounced minimum near $B=0$.

\begin{figure}[ht!]
\epsfysize=60mm \epsfbox[24 83 548 612]{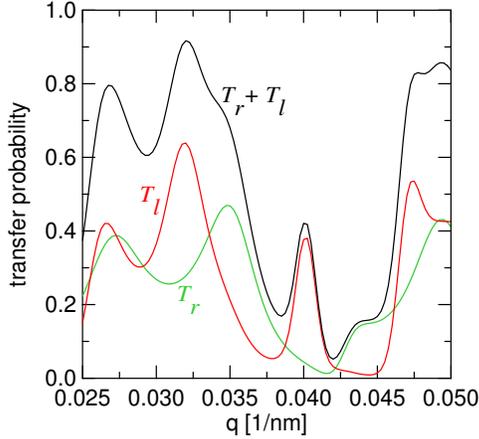} \\
\caption{Electron transfer probability to the left $T_l$ and right $T_r$ leads for the ring
with a repulsive defect of height 3 meV and $B=0$
in function of the wave vector $q$.}
  \label{g3k}
\end{figure}

\begin{figure*}[ht!]
\begin{tabular}{llll}
a) & b) &c) & d)\\
\epsfysize=37mm \epsfbox[24 109 540 610]{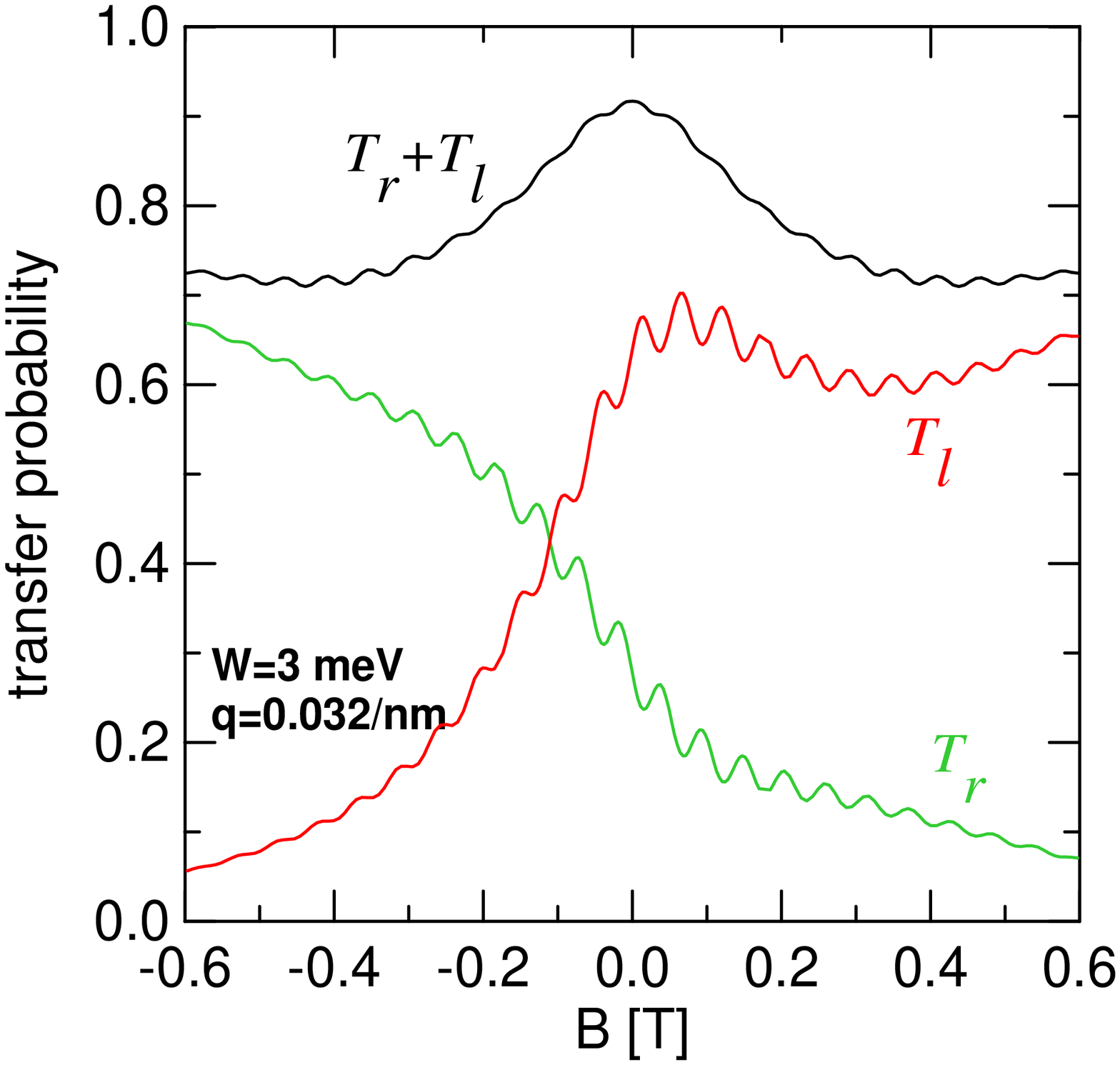} &
\epsfysize=37mm \epsfbox[24 109 540 610]{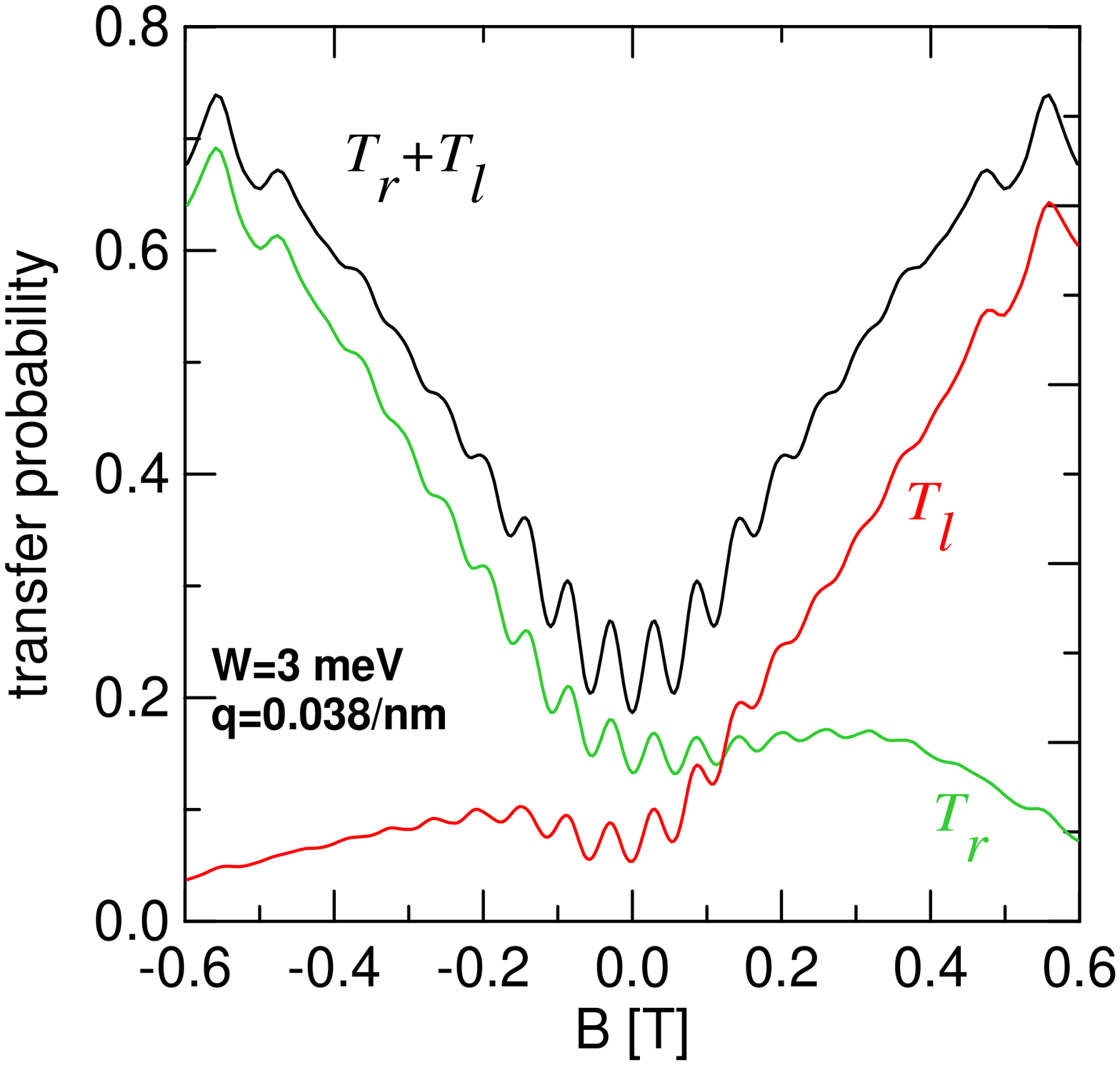} &
\epsfysize=37mm \epsfbox[24 109 540 610]{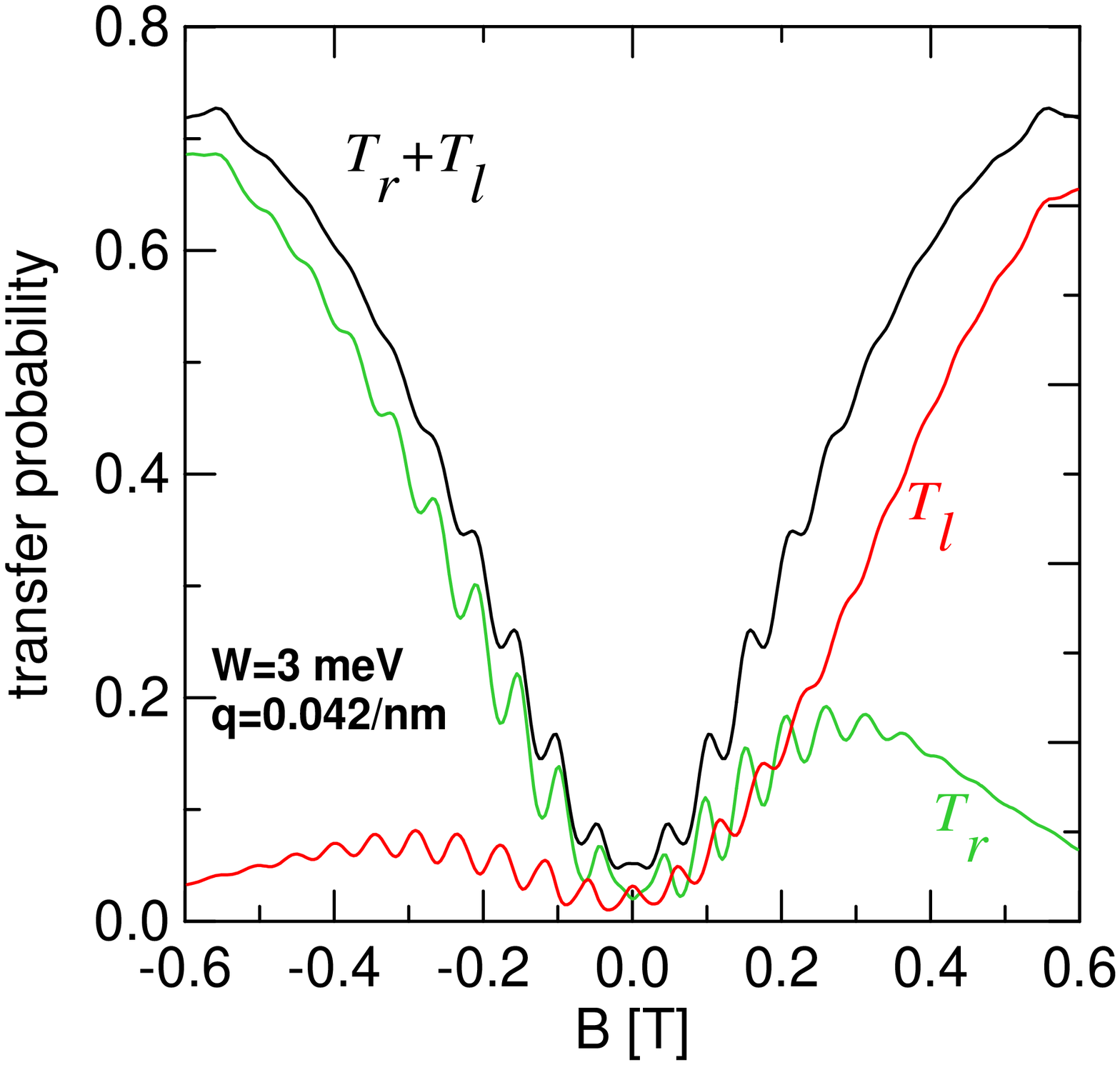} &
\epsfysize=37mm \epsfbox[24 109 540 610]{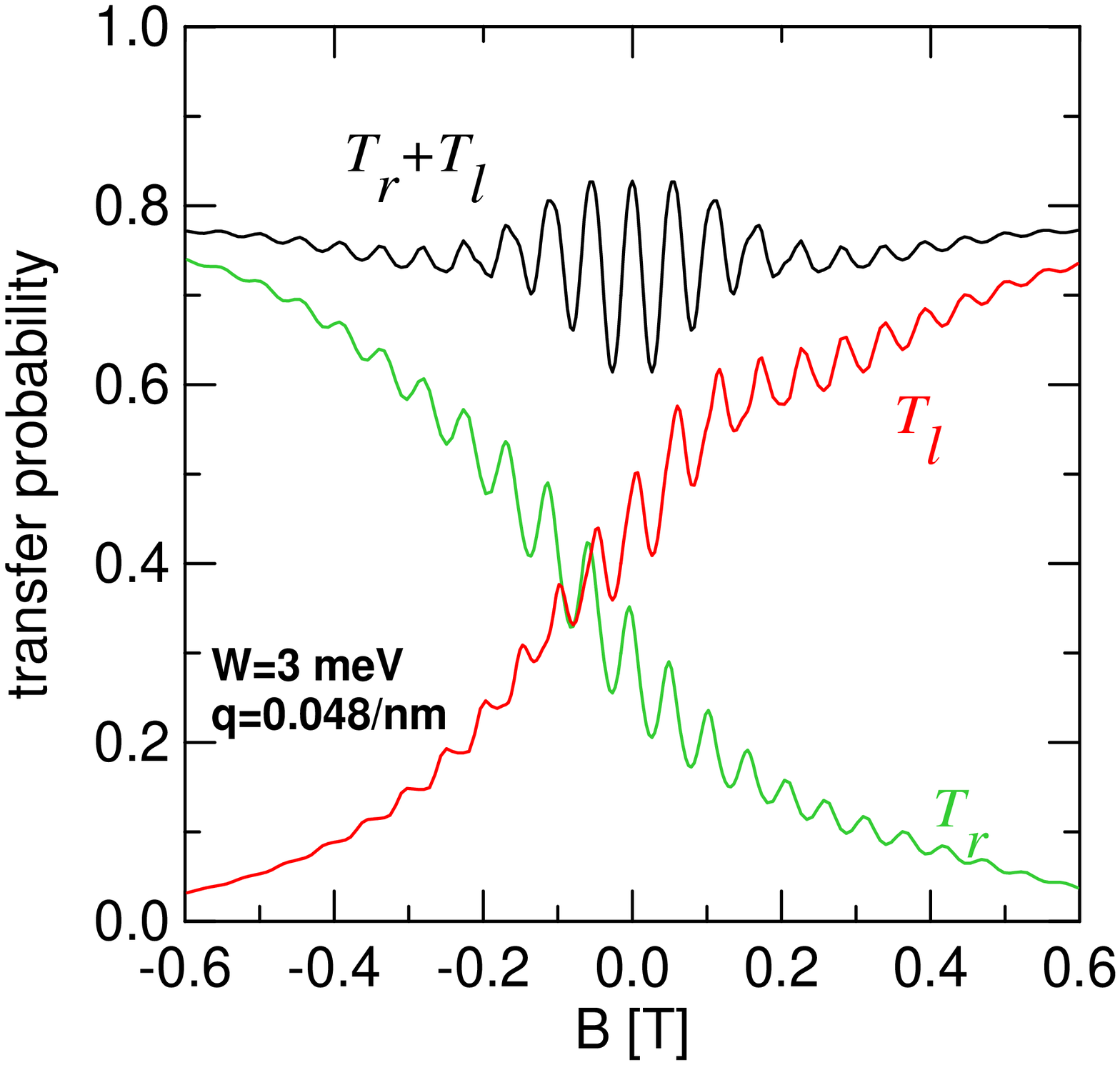} \\
\end{tabular}
\caption{Electron transfer probability for the repulsive defect $W=3$ meV for
average wave vector values $q=0.032$/nm (a), $q=0.038$/nm (b), $q=0.042$/nm (c) and $q=0.048$/nm (d) in function of the magnetic field.}
  \label{wq}
\end{figure*}

In order to establish which of the above features are independent of the wave vector we plotted in Fig. \ref{g3k} the transfer probabilities
for $B=0$ in function of $q$ for fixed value of the height of the defect $W=3$ meV, and in Fig. \ref{wq} the transfer probabilities
in function of the magnetic field for several fixed values of $q$.
In Fig. \ref{g3k} we see that the defect present in the left arm of the ring does not necessarily imply $T_l<T_r$, for $B=0$.
Results of Fig. \ref{wq} (a) and (d) were calculated for local maxima of $T_l$. The crossings between $T_l$ and $T_r$ appears here
for the negative magnetic field instead of the positive $B$ as in Fig. \ref{wv}(e-g). The overall transfer probability has a maximum
near $B=0$ for $q=0.032$ / nm, and for $q=0.048$ / nm the value of $T$ oscillates around an average constant value.
For $q=0.042$/nm [Fig. \ref{wq}(c)] -- near the minimum of $T(q)$ (Fig. \ref{g3k}) $T_l$ and $T_r$ have similar values on a longer
range of $B$ between $0$ and $0.2$ T. The presented results indicate that of the above list of features only i), iii) and v)
are characteristic to the ring with a strong scattering repulsive defect, and the rest is $q$--dependent.
The Fermi wave vector in the experimental samples is determined by the
density of the two-dimensional electron gas in the electron reservoirs. In principle the electron density is fixed at the sample formation
stage by the dopants concentration  within the AlGaAs barrier, but  the wave vector should be at least to an extent tunable by
the voltages applied to the electrodes in a gated sample.

In order to comment on the location of the defect that we consider here, we indicate that a defect in between
the input and the right output lead produces the transition spectra with inverted value of the magnetic field $(B\rightarrow -B)$
and a defect placed exactly in the center of the arm between the output leads produces symmetric spectra only with oscillations amplitude
that is reduced for $W>0$. For $B=0$ the exact position of the defect within the same section of the ring influences the position
of the transfer probability peaks on the wave vector scale, but otherwise no qualitative difference is found in the transfer characteristics
as function of the external magnetic field.

\section{Summary and Conclusions}
We have studied the electron transport through a three-terminal quantum ring containing elastic scatterers using a time-dependent approach.
The presented study indicates that elastic scattering may be a significant reason
of the low amplitude of Aharonov-Bohm oscillations simultaneously explaining the low-field asymmetry
of the conductance to both the output leads as observed in a recent experiment.
Low visibility of the oscillations introduced by the elastic scattering is not due to the phase randomization
but to a hindered circulation of the electron around the ring.
Both decoherence and intersubband scattering \cite{chaves} that were neglected in the presented study
should also reduce the visibility of the Aharonov-Bohm oscillations but by themselves they cannot introduce the strong asymmetry effects
that are distinct in the experimental results.
We also found that the potential defect present within the ring affects the properties of the transmission probability only in the low magnetic field. The high magnetic field limit is left unchanged:
the Aharonov-Bohm oscillations
are reduced by the electron injection imbalance due to the magnetic forces and  the conductance of one of the leads increases at the expense of the other
as in the case of a clean quantum ring.

{\bf Acknowledgements} This work was supported by the
AGH UST project 11.11.220.01 "Basic and applied research in nuclear
and solid state physics". Calculations were performed in
ACK\---CY\-F\-RO\-NET\---AGH on the RackServer Zeus.


\begin{thebibliography}{00}
\bibitem{but} M. B\"uttiker, Y. Imry, and M. Y. Azbel, Phys. Rev. A {\bf 30}, 1982
(1984).
\bibitem{tip} G. Timp, A.M. Chang, J.E. Cunningham, T.Y. Chang, P. Mankiewich,
R. Behringer, and R. E. Howard, Phys. Rev. Lett. {\bf 58}, 2814 (1987).
\bibitem{firer} A. Fuhrer, S. L\"uscher, T. Ihn, T. Heinzel, K. Ensslin, W.
Wegscheider, and M. Bichler, Nature (London) {\bf 413}, 822 (2001).
\bibitem{keyser} U.F. Keyser, C. F\"uhner, S. Borck, R. J. Haug, M. Bichler, G.
Abstreiter, and W. Wegscheider, Phys. Rev. Lett. 90, 196601
(2003).
\bibitem{muhle} A. M\"uhle, W. Wegscheider and R.J. Haug, Appl. Phys. Lett. {\bf 91}, 133116 (2007).
\bibitem{gustav} S. Gustavsson, R. Leturcq R, M. Studer, T. Ihn T, K. Ensslin, D.C. Driscoll, and A.C. Gossard A. C., Nano Lett., {\bf 8}, 2547 (2008).
\bibitem{sgs} F. Martins, B. Hackens, M.G. Pala, T. Ouisse, H. Sellier, X. Wallart, S. Bollaert, A. Cappy, J. Chevrier, V. Bayot, and S. Huant,
Phys. Rev. Lett. {\bf 99}, 136807 (2007).
\bibitem{wlochy} E. Strambini, V. Piazza, G. Biasiol, L. Sorba, and F. Beltram, Phys. Rev. B {\bf 79}, 195443 (2009).
\bibitem{time} B. Szafran and F.M. Peeters, Phys. Rev. B {\bf 72}, 165301 (2005).
\bibitem{epl} B. Szafran and F.M. Peeters, Europhys. Lett. {\bf 70}, 810 (2005).
\bibitem{dec} M. Ferrier, L. Angers, A.C.H. Rowe, S. Gu\'eron, H. Bouchiat, C. Texier, G. Montambaux, and D. Mailly,
Phys. Rev. Lett. {\bf 93}, 246804 (2004).
\bibitem{onsager} M. B\"uttiker, Phys. Rev. Lett. {\bf 57}, 1761 (1986).
\bibitem{acshift} T. Bergsten, T. Kobayashi, Y. Sekine, and J. Nitta, Phys. Rev.
Lett. {\bf 97}, 196803 (2006). 
\bibitem{peeters} P. F\"oldi, O. Kalman, M. G. Benedict, and F.M. Peeters, Nano Lett. {\bf 8}, 2556 (2008). 
\bibitem{lt} R. Leturcq, L. Schmid, K. Ensslin, Y. Meir, D.C. Driscoll, and A. C. Gossard, Phys. Rev. Lett. {\bf 95}, 126603 (2005). 
\bibitem{bsp} P. F\"oldi, O. Kalman, M. G. Benedict, and F.M. Peeters, Phys. Rev. B {\bf 73}, 155325 (2006);
O. Kalman, P. F\"oldi, M.G. Benedict, F.M. Peeters, Physica E {\bf 40}, 567 (2007). 
\bibitem{kali} R. Kalina, B. Szafran, S. Bednarek, and F.M. Peeters, Phys. Rev. Lett. {\bf 102}, 066807 (2009).
\bibitem{bqd} F.M. Peeters, in {\it Science and Engineering of One- and Zero-Dimensional Semiconductors}, edited by S.P. Beaumont and C.M. Sotomajor Torres
 (Plenum, New York, 1990), p. 107.
\bibitem{pml} C. Farell and U. Leonhardt, J. Opt. B: Quantum Semiclass. Opt. {\bf 7}, 1 (2005).
\bibitem{tns} A. Arnold, M. Ehrhardt, and I. Sofronov, Comm. Math. Sci. {\bf 1}, 501  (2003).
\bibitem{chaves} A. Chaves, G.A. Farias, F.M. Peeters, and B. Szafran, Phys. Rev. B {\bf 80} - in print (BG11520)  
\bibitem{buttikernowy} M. B\"uttiker, Y. Imry, and M. Ya. Azbel, Phys. Rev. A {\bf 30}, 1982 (1984). 
\bibitem{sols} F. Sols, M. Macucci, U. Ravaioli, and K. Hess, Appl. Phys. Lett. {\bf 54}, 350 (1989).
\bibitem{askar} A. Askar and A.C. Cakmak, J. Chem. Phys. {\bf 68}, 2794 (1978).
\end{thebibliography}
\end{document}